\documentclass{ws-procs9x6}

\begin{document}

\title{Casimir-Polder forces between two accelerating atoms and the Unruh effect}

\author{Jamir Marino* and Roberto Passante**}

\address{Dipartimento di Scienze Fisiche ed Astronomiche \\dell'Universit\`{a} degli Studi di Palermo and CNISM,
Via Archirafi 36, \\
I-90123 Palermo, Italy\\
* E-mail: jamirmarino@gmail.com\\
** E-mail: roberto.passante@fisica.unipa.it}

\begin{abstract}
The Casimir-Polder force between two atoms with equal uniform acceleration and separated by a constant distance $R$  is considered. We show that, in the low-acceleration limit, while the near-zone $R^{-6}$ behavior of the interatomic interaction energy is not changed by the acceleration of the atoms, the far-zone interaction energy decreases as $R^{-5}$ instead of the well-known $R^{-7}$ behavior for inertial atoms.  Possibility of an indirect detection of the Unruh effect through measurements of the Casimir-Polder force between the two accelerating atoms is also suggested. We also consider a heuristic model for calculating the Casimir-Polder potential energy between the two atoms in the high-acceleration limit.
\end{abstract}

\keywords{Unruh effect; Casimir-Polder forces.}

\bodymatter

\section{Introduction}

Uniformly accelerated observers in Minkowski spacetime, i.e. linearly accelerated observers with constant proper
acceleration, called \emph{Rindler observers}, associate a
thermal bath of Rindler particles to the scalar-field no-particle state of inertial observers, the Minkowski vacuum. This effect was discovered by W.G. Unruh \cite{Unruh} and has played an important role in our understanding that the particle content
of a field theory is observer-dependent \cite{Fulling}. Detection of the Unruh effect however requires very high accelerations, $a\simeq 10^{20} m/s^2$, and many proposals have been made in order to get a direct measurement of the Unruh effect, analyzing for example electrons in particle accelerators \cite{Leinas} or atoms accelerating in microwave cavities \cite{Bely} .

 A change in the particle content of the vacuum state can in principle produce a change in any physical phenomena directly related to the vacuum fluctuations, such as the Lamb Shift of atomic levels or Casimir forces. Recently, the corrections to the Casimir-Polder force between an accelerating atom and a conducting plate due to the Unruh effect  have been calculated, both in the scalar- and in the electromagnetic-field case \cite{Rizzuto1,Rizzuto2},  using an appropriate extension of an approach which allows separation of vacuum fluctuations and radiation reaction contributions \cite{DDC} . These works have shown that modifications to the atom-plate Casimir-Polder force are relevant only for accelerations of the order of $10^{20} m/s^2$, confirming the necessity of high accelerations for a direct detection of the Unruh effect. Also, it has been shown that in the high-acceleration regime, atoms are spontaneously excited, absorbing a quantum from the Unruh bath \cite{Rizzuto1}; in this way the atom-wall force becomes reminiscent of that between a static excited atom and a plate.

In this paper, we consider the case of the Casimir-Polder force between two atoms with the same uniform acceleration and separated by a constant distance $R$. Both the low-acceleration and the high-acceleration limits ($a\ll\omega_0c$ and $a\gg\omega_0c$, respectively) are discussed, using an approach first developed by Goedecke and Wood \cite{Goedecke} for the Casimir-Polder force between two atoms at finite temperature. We show that in the low-acceleration regime the Unruh effect changes the far-zone law of the interatomic potential energy from the $R^{-7}$ dependence characteristic of two inertial atoms, to a $R^{-5}$ dependence; in the near zone we find a correction to the potential energy due to the acceleration scaling as $R^{-6}$, as for inertial atoms. We also argue about the possibility of an indirect measurement of the Unruh effect, exploiting this change in the behavior of the interatomic potential. In the high-acceleration regime, we obtain an interatomic potential energy reminiscent of that between two excited-state atoms.

\section{The Casimir-Polder force between two accelerating atoms}

When a neutral polarizable atom uniformly accelerates in the electromagnetic vacuum, a radiative shift of its energy levels occurs due to the presence of the electromagnetic Unruh bath.
These energy shifts are known in the literature \cite{Passanteunruh}, and they provide the following expression for the average field energy of the Unruh bath in each mode divided by $\hbar \omega_k$,
\begin{equation}\label{numero modi}
\langle n(\omega ) \rangle_a=\frac 12 \Big(1+\frac{a^2}{c^2\omega^2}\Big)\Big(1+\frac{2}{e^{2\pi c\omega/a}-1}\Big).
\end{equation}
In \eqref{numero modi} there is a non-thermal contribution proportional to $a^2/c^2\omega^2$, which in the case of the electromagnetic field breaks the equivalence between a thermal bath and the Unruh spectrum, which is valid in scalar field theories \cite{Passanteunruh}.

We now consider a pair of neutral polarizable atoms, both accelerating with the same uniform acceleration $a$ in a direction orthogonal to their separation, measured from the laboratory reference frame. Because the two atoms have the same acceleration, they see the same electromagnetic Unruh bath, and in order to estimate the Casimir-Polder force between them we use a technique developed by Goedecke and Wood \cite{Goedecke} for the interatomic force at temperature $T$. Following this approach, and taking into account that the two atoms see the same Unruh bath, we consider the expression for the Casimir-Polder potential for two atoms in the vacuum
\begin{equation}\label{cp}
V(R)=-\frac{2\hbar c}{\pi R^2}\textmd{Im}\int_0^\infty dkk^4\Big(\frac{1}{2}\Big)e^{i2kR}U(kR)\alpha^2(k)
\end{equation}
($\alpha (k)$ is the dynamical atomic polarizability) with
\begin{equation}
U(x)=1+5x^{-2}+3x^{-4}+i\left( 2x^{-1} +6x^{-3} \right) ,
\end{equation}
and we substitute in \eqref{cp} the average energy per mode in the vacuum state divided by $\hbar \omega_k$ (which is equal to $\frac 12$), with the analogous quantity for the Unruh bath of Rindler particles, given by \eqref{numero modi}.
This gives the following expression for the Casimir-Polder potential between two accelerating atoms
\begin{equation}\label{tre}
V(R)=-\frac{2\hbar c}{\pi R^2}\textmd{Im}\int_0^\infty dkk^4\langle n(\omega ) \rangle_ae^{i2kR}U(kR)\alpha^2(k),
\end{equation}
in which we have assumed that the acceleration is so small that no spontaneous excitation occurs in the two atoms, that is $a\ll\omega_0c$.
We can develop the calculations in \eqref{tre} similarly as in Ref. \cite{Goedecke} by a rotation in the first quadrant of the complex $k$-plane and separating the contributions from the principal part of the integral from the contributions from the poles of the integrand. If we go straightforwardly with this method, we get two different expressions for the potential energy in two different zones, $\frac{aR}{c^2}\ll1$ and $\frac{aR}{c^2}\gg1$, in a similar way to the case of the thermal Casimir-Polder potential energy \cite{Goedecke}.

For $\frac{aR}{c^2}\ll1$, we find in the \emph{near zone} ($R\ll c/\omega_0$) the following expression for the interatomic potential energy
\begin{equation}\begin{split}
V(R)=&-\frac{2}{3R^6}\sum_{r,s}\frac{\mu_{r0}^2\mu_{s0}^2}{E_{r0}+E_{s0}}\\
-&\frac{4}{3}a^2\frac{\hbar}{\pi}\frac{1}{c^3}\frac{1}{R^6}\sum_{r,s}E_{r0}E_{s0}\mu_{r0}^2\mu_{s0}^2f(E_{r0},E_{s0})
\end{split}\end{equation}
where $\mu_{r0}$ are matrix elements of the atomic dipole moment and $f(E_{r0},E_{s0})$ is a function of the atomic transition frequencies.
This shows that the Unruh effect introduces in the \emph{near zone} only a correction proportional to the square of the acceleration of the two atoms, which scales with the same power law of the inertial Casimir-Polder potential energy.
In the \emph{far zone} ($R\gg c/\omega_0$) we instead get a correction to the static Casimir-Polder force, which scales as $R^{-5}$,
\begin{equation}
\label{fz}
V(R) = -\frac{23\hbar c}{4}\frac{\alpha^2(0)}{R^7}-\frac{\hbar a^2}{4\pi c^3}\frac{\alpha^2(0)}{R^5}.
\end{equation}
The new acceleration-dependent term introduced here by the Unruh effect changes the dependence of the potential from the interatomic distance $R$, and in principle it could be used  to make an indirect measurement of the Unruh effect, for example by measuring some macroscopic quantity of an ensemble of atoms sensitive to the form of the potential. This new term gives an attractive contribution to the interatomic force, increasing its strength.

For $aR/c^2 \gg 1$ we find in both the near and far zones an expression of the potential energy that depends on the interatomic distance as $R^{-6}$,
\begin{equation}
\label{ht}
V(R) = -\frac{6\hbar a\alpha^2(0)}{\pi R^6c}\Big(\frac{1}{4}+\frac{\pi^2}{12}\Big).
\end{equation}
This result is similar to that obtained by Goedecke and Wood \cite{Goedecke} for the Casimir-Polder interaction at finite temperature in the limit $2\pi k_BTR/ \hbar c \gg1$.

The results \eqref{fz} and \eqref{ht}  are obtained in the hypothesis  of small accelerations, such that the spontaneous excitation probability of the atom be negligible. For $a \gg \omega_0 c$, spontaneous excitation must be taken into account \cite{AuMu2}. In such a case, we can obtain the interatomic interaction energy by substituting the average field energy of the Unruh bath in each mode, divided by $\hbar \omega_k$, \eqref{numero modi} in the expression for the Casimir-Potential between two excited atoms in terms of vacuum field correlations as obtained by Power and Thirunamachandran \cite{Power}.

The most relevant contribution in this case is given by the Rindler particles resonant with the atomic transition frequency $ck_A$. For $a\gg\omega_0c$ and $k\simeq k_A$, from \eqref{numero modi} we get
\begin{equation}
\langle n(\omega_A)\rangle_a\simeq\frac{a^3}{2\pi c^6k_A^3}.
\end{equation}
Thus, with our method, the final expression for the interatomic potential in the high-acceleration regime is
\begin{equation}
\Delta E=-\frac{2}{3}\frac{\mu_A^2 \alpha_B(k_A)a^3k_A}{\pi R^2c^6}\Big[1+\frac{1}{k_A^2R^2}+\frac{3}{k_A^4R^4}\Big],
\end{equation}
where $\mu_A$ is the dipole moment of atom A. This expression shows a $R^{-6}$ behaviour in the near zone, and a $R^{-2}$ behavior in the far zone, reminiscent of the interatomic potential energy between two excited static atoms \cite{Power} and coherent with the results for the atom-plate Casimir-Polder interaction \cite{Rizzuto1}.

Finally, a few words are necessary on the limits that our approximations imply. The non-relativistic approach, $v \ll c$,  implies a constraint on the timescale in which our calculations are valid, that is  $\tau \ll c/a$; on the other hand, this timescale must be greater than the characteristic time of atomic transitions $\omega_0^{-1}$,  in order to let Casimir-Polder forces between the two atoms to be established.
The limits above  are the same assumed also in previous works on the role of the Unruh effect in atomic systems \cite{AuMu}.

\section{Conclusions}
We have studied the Casimir-Polder force between two uniformly accelerating atoms both in the low- ad high-acceleration regime. In the low acceleration limit, $a\ll\omega_0c$, we have obtained the Unruh correction to the interatomic Casimir-Polder force. Our results show that in the \emph{near zone} the Unruh electromagnetic bath only gives a correction to the $R^{-6}$ law of the interatomic interaction energy, while in the \emph{far zone} the Unruh correction changes the $R$-dependence of the potential energy, adding an attractive acceleration-dependent $R^{-5}$ contribution  to the inertial $R^{-7}$ term.
In the high-acceleration limit, we find that the Casimir-Polder potential energy shows a $R^{-2}$ dependence in the far zone, related to the possibility of spontaneous excitation of the atoms.

\section*{Acknowledgements}
J.M. acknowledges financial support by the European Science Foundation
(ESF) within the activity `New Trends and Applications of the Casimir
Effect' (www.casimir-network.com).
Partial financial support from Ministero dell'Istruzione, dell'Universit\`{a} e della Ricerca and by Comitato Regionale di Ricerche
Nucleari e di Struttura della Materia is also acknowledged.

\bibliographystyle{ws-procs9x6}

\end{document}